\documentclass[runningheads]{llncs}
\usepackage{graphicx, multirow, slashbox}
\usepackage{comment}

\begin{document}
\title{MetaHistoSeg: A Python Framework for Meta Learning in Histopathology Image Segmentation}

\author{Zheng Yuan \and Andre Esteva \and Ran Xu }

\institute{Salesforce Research, Palo Alto CA 94301, USA \email{zyuan@salesforce.com} }
\maketitle              

\begin{abstract}
Few-shot learning is a standard practice in most deep learning based histopathology image segmentation, given the relatively low number of digitized slides that are generally available. 
While many models have been developed for domain specific histopathology image segmentation, cross-domain generalization remains a key challenge for properly validating models. Here, tooling and datasets to benchmark model performance \emph{across} histopathological domains are lacking. To address this limitation, we introduce MetaHistoSeg -- a Python framework that implements unique scenarios in both meta learning and instance based transfer learning. Designed for easy extension to customized datasets and task sampling schemes, the framework empowers researchers with the ability of rapid model design and experimentation.    
We also curate a histopathology meta dataset - a benchmark dataset for training and validating models on out-of-distribution performance across a range of cancer types. In experiments we showcase the usage of MetaHistoSeg with the meta dataset and find that both meta-learning and instance based transfer learning deliver comparable results on average, but in some cases tasks can greatly benefit from one over the other.  


\end{abstract}

\section{Introduction}


For cancer diagnosis and therapeutic decision-making, deep learning has been successfully applied in segmenting a variety of levels of histological structures: from nuclei boundaries~\cite{nuclei} to epithelial and stromal tissues~\cite{epithelial vs. stromal regions}, to glands~\cite{gland1,gland2} across various organs. 
It's generalizability that makes it effective across a wide variety of cancers and other diseases. 
Admittedly, the success relies largely on the abundance of datasets with pixel level segmentation labels \cite{gleason2019,breastQ,MoNuSeg,glandsegmentation,DigestPath}. 

Few-shot learning is of particular importance to medicine. 
Whereas traditional computer vision benchmarks may contain millions of data points, histopathology typically contains hundreds to thousands. 
Yet in histopathology images, different cancers often share similar visual components. For instance, adenocarcinomas, which occur in glandular epithelial tissue, contain similar morphological structure across many organs where they can arise~\cite{cancerfact}. Thus models that distill transferable histopathological features from one cancer can potentially transfer this knowledge to other cancers.
To utilize different histopathology datasets collectively, benchmarks and tooling that enable effective learning across domains are strongly desired to support more accurate, and more generalizable models across cancers.

The key question is how to formulate the learning-across-task setup for histopathology segmentation? Naturally meta-learning~\cite{metalearning1,metalearning2,metalearning3} is the best reference as for its precise effectiveness to handle limited data availability. It is widely used in few shot classification with a canonical setup: a task of K-way-N-shot classification is created on the fly by sampling K classes out of a large class pool followed by sampling N instances from each of the K classes. Then a deep neural network is trained by feeding batches of these artificial tasks. Eventually during inference the whole network is shared with new tasks (composed by K classes never seen during training) for refinement.

While this setup is ubiquitous in meta classification, we find that it is difficult to extend to the meta segmentation problem. First, a task of segmenting histopathology images should justify medical validity (e.g. cancer diagnosis) before even created. One cannot generate factitious tasks by randomly combining K layers of pixels based on their mask label, as oppose to the routine in meta classification. 
For example, based on a well-defined Gleason grading system, a prostate cancer histopathology image usually requires to be classified into 6 segments for each pixel. Meanwhile for another histopathology image in nuclei segmentation, researchers in general need to classify each pixel as either nuclei or others. Notwithstanding each case exhibits a valid medical task in its own right, criss-crossing them just as in the canonical setup to form a new task is not medically sound.
Moreover, the underlying assumption of meta classification is that shared knowledge must exist across any K-way classification tasks. It is generalizable among tasks by a composite of any K classes, as long as the number of classes involved is K. Generally, we will not observe this "symmetrical" composite in segmentation task space. In the same example, the first task is to segment 6 classes pixel wise whereas the other is to segment 2 classes. Therefore, the knowledge sharing mechanism in the deep neural also needs to be adjusted to reflect this asymmetry.


In this paper, we introduce a Python framework MetaHistoSeg to facilitate the proper formulation and experimentation of meta learning methodology in histopathology image segmentation. We also curate a histopathology segmentation meta-dataset as the exemplar segmentation task pool to showcase the usability of MetaHistoSeg. 
To ensure the medical validity of the meta dataset, we build it from existing open-source datasets that are (1) rigorously screened by world-wide medical challenges and (2) well-annotated and ready for ML use.

MetaHistoSeg offers three utility modules that cover the unique scenarios in meta learning based histopathology segmentation from end to end:

1) Data processing functions that normalize each unique dataset pertaining to each medical task into a unified data format.

2) Task level sampling functions (the cornerstone of the meta learning formulation in segmentation) for batch generation and instance level sampling functions provided as a baseline.

3) Pre-implemented task-specific heads that are designed to tail customized backbone to handle the asymmetry of tasks in a batch, with multi-GPU support. 

We open-source both MetaHistoSeg and the meta-dataset for broader use by the community. The clear structure in MetaHistoSeg and the accompanying usage examples allow researcher to easily extend its utility to customized datasets for new tasks and customized sampling methods for creating task-level batches. Just as importantly, multi-GPU support is a must in histopathology segmentation since a task level batch consists of fair number of image instances, each of which is usually in high resolution. 
We also benchmark the performance of meta learning based segmentation as compared with the instance based transfer learning as a baseline. Experiments show that both meta-learning  and instance based transfer learning and deliver comparable results on average, but in some cases tasks can greatly benefit from one over the other.

\section{MetaHistoSeg framework}

MetaHistoSeg offers three utilities: task dataset preprocessing, task or instance level batch sampling, task-specific deep neural network head implementation. 

\subsection{Histopathology task dataset preprocessing}

MetaHistoSeg provides preprocessing utility functions to unify the heterogeneity of independent data sources with a standard format. Here we curate a meta histopathology dataset to showcase how knowledge transfer is possible via meta learning among different segmentations tasks. Following the tasks in the dataset as examples, users can easily create and experiment with new tasks from customized datasets.  

The meta-dataset integrates a large number of histopathology images that come from a wide variety of cancer types and anatomical sites.
The contextual information of each data source, the preprocessing method and the meta information of their data constituents are detailed as follows. 

\begin{itemize}
  \item \textit{Gleason2019}: a dataset with pixel-level Gleason scores for each stained prostate cancer histopathology image sample. Each sample has up to six manual annotations from six pathologists. During preprocessing, we use the image analysis toolkit SimpleITK\cite{staple} to consolidate multiple label sources into a single ground truth. The dataset contains 244 image samples with resolution of 5120x5120 and each pixel belongs to one of 6 Gleason grade grades. The data source was a challenge\cite{gleason2019} hosted in MICCAI 2019 Conference.
  
  \item \textit{BreastPathQ}: a dataset of patches containing lymphocytes, malignant epithelial and normal epithelial cell nuclei label. This is an auxiliary dataset in the Cancer Cellularity Challenge 2019\cite{breastQ} as part of the 2019 SPIE Medical Imaging Conference where the original task is to evaluate patch as a single score. In our context, we use the dataset for segmentation. Since the annotations only contain the centroid of each cell nuclei, we generate the segmentation mask by assuming each cell is a circle with a fixed radius. The dataset contains 154 samples and each pixel belongs to one of 4 classes. 

  \item \textit{MoNuSeg}: a dataset of pixel-level nuclei boundary annotations on histopathology images from multiple organs, including breast, kidney, liver, prostate, bladder, colon and stomach. This dataset comes from the nuclei segmentation challenge\cite{MoNuSeg} as an official satellite event in MICAII 2018. It contains 30 samples and each label has 2 classes.
    
  \item \textit{Glandsegmenatation}: a dataset of pixel-level gland boundary annotations on colorectal histopathology images. This data source comes from the gland segmentation challenge\cite{glandsegmentation} in MICAII 2015. The dataset contains 161 samples and each label has 2 classes.

  \item \textit{DigestPath}: a dataset of colon histology images with pixel-level colonoscopy lesion annotations. The data source\cite{DigestPath} is part of MICCAI 2019 Grand Pathology Challenge. It contains 250 samples and each pixel belongs to one of the 2 classes. Although the original challenge contains both Signet ring cell detection and Colonoscopy tissue segmentation task, we only consider the latter in our context for the obvious reason.
\end{itemize}

\subsection{Task and instance level batch sampling}

MetaHistoSeg implements this core data pipeline of meta learning. It abstracts task level batch creation as a dataloader class \textit{episode\_loader}. Since \textit{episode\_loader} essentially unrolls the entire task space, researchers can customize their sampling algorithm just by specifying a probability distribution function.
This enables users to quickly switch between training frameworks, empowering them to focus on model design and experimentation rather than building data pipelines. It also encapsulates instance level batch creation in dataloader class \textit{batch\_loader} as a baseline. The sampling schemes are as follows, 
\begin{itemize}
  \item Task level sampling: we sample a task indexed by its data source and then sample instances given the task to form an episode. Then it is split into support and query set. Here a batch is composed of several such episodes.
  \item Instance level sampling: we first mix up instances from different data sources as a pool and sample instances directly. Noting that data source imbalance can be a problem here, we dynamically truncate each data source to the same size before mixing up. We refresh the random truncation in each epoch. 
\end{itemize}

\begin{figure}
\begin{centering}
\includegraphics[width=320pt]{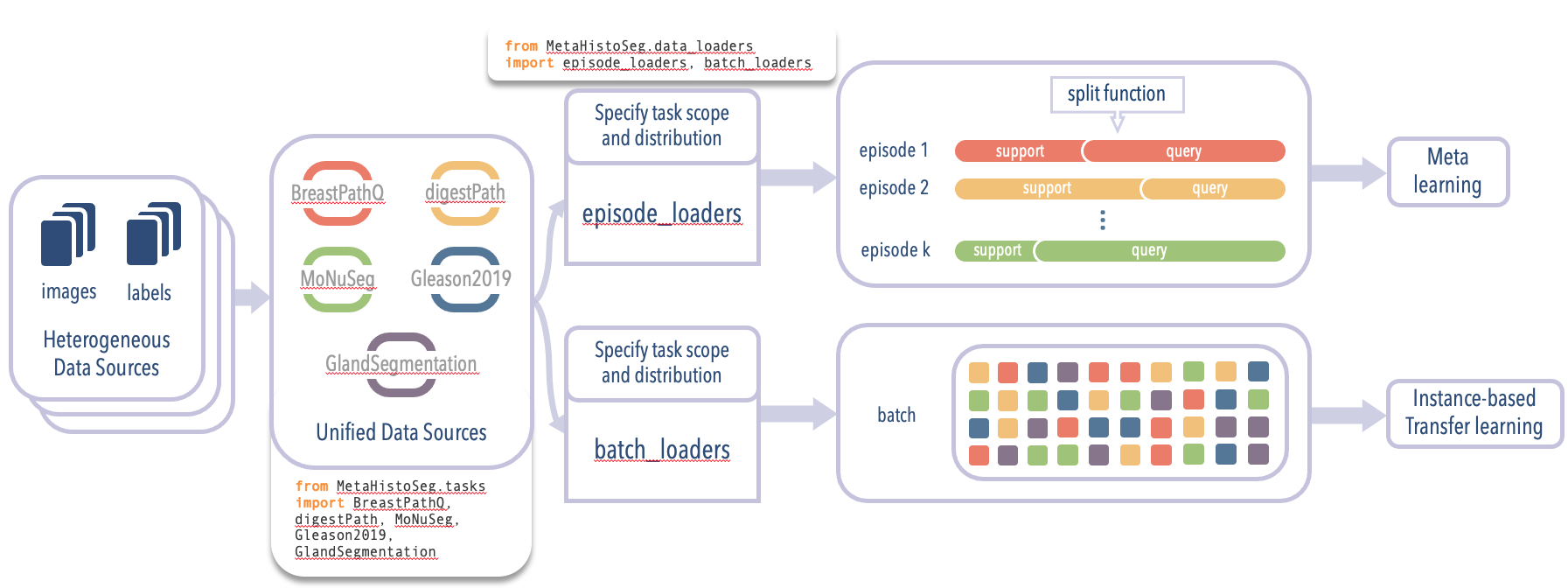}
\caption{MetaHistoSeg diagram: utility functions enable fast construction of data pipelines for meta-learning and instance based learning on the meta-dataset.} \label{fig:datapipeline}
\end{centering}
\end{figure}

Fig.~\ref{fig:datapipeline} shows how the preprocessing and batch sampling functions in MetaHistoSeg can be used to construct the data pipelines. Each data source is color coded. In meta learning setting, a batch is organized as episodes, each of which comes from the same data source. In instance based learning, a batch is organized as instances, which comes from mixed up data sources.

\subsection{Task-specific heads and multi-GPU support}

Since the tasks sampled in a batch usually predict different number of segments, we pre-implement the last layer of a neural network as task-specific heads and route the samples of a task only to its own head during forward propagation (FP). This feature frees researchers from handling task asymmetry in meta segmentation. Meanwhile, note that the default multi-GPU support in pytorch (nn.DataParallel) requires a single copy of network weights. This conflicts with the meta learning scenario, where two copies of weight parameters are involved in its bi-level optimization. Thus we re-implement multi-GPU FP process. 

\section{Experiments}
We use MetaHistoSeg to benchmark MAML\cite{metalearning1} on the histopathology image segmentation meta dataset and compare it with instance based transfer learning as a baseline. 
For each data source in the meta-dataset, we fix it as a test task and train a model using some subset of the remaining data sources, using both MAML and instance based transfer learning.

\subsection{Implementation Details}
For data augmentation, we resize an input image with a random scale factor from 0.8 to 1.2, followed by random color jittering (with 0.2 variation on brightness, contrast and 0.1 variation on hue and saturation),  horizontal and vertical flipping (0.5 chance) and rotation (a random degree from -15 to 15). The augmented image is ultimately cropped to 768x768 before feeding into the neural networks. 

During training, we use 4 Nvidia Titan GPU (16G memory each) simultaneously. This GPU memory capacity dictates the maximum batch size as 4 episodes with each consisting of 16 image samples. During meta learning, each episode is further split into a support set of size 8 and a query set of size 8. When forming a batch, we use MetaHistoSeg.episode\_loader to sample data in a bi-level fashion: first among data sources then among instances.

In both methods we choose U-Net\cite{unet} as the backbone model given its effectiveness at medical image segmentation tasks. Training is performed with an Adam optimizer and a learning rate of 0.0001 for both methods. For MAML, we adapt once with a step size of 0.01 in the inner loop optimization. The maximum training iteration is set to $300000$ for both settings.

We use the mean Intersection Over Union (mIoU) between predicted segmentation and ground truth as our performance metric: 
\begin{equation}
\mbox{mIoU} = \frac{1}{N} \sum_{i}{\frac{P_i \cap T_i  }{P_i \cup T_i }}
\end{equation}
where $P_i$ and $T_i$ are the predicted and ground truth pixels for class $i$, respectively, across all images in evaluation, and $N$ is the number of classes.

\subsection{Results}

Table \ref{table:results} summarizes the mIoU scores for both methods where each data source is treated as a new task, and models are trained on some subset of the remaining data sources. We enumerate over the other data sources as well as their combination to form five different training sets - the five columns in the table.

\begin{table}[]
\caption{mIoU performance on each new task (row) refined from pretrained models with different predecessor tasks (column)}
\label{table:results}
\centering
\scalebox{0.75}{
\begin{tabular}{|l|l|l|l|l|l|l|l|l|l|l|}
\hline
\multicolumn{1}{|c|}{\multirow{2}{*}{\backslashbox{new task}{training tasks}}} & \multicolumn{2}{l|}{All others} & \multicolumn{2}{l|}{BreastPathQ}                  & \multicolumn{2}{l|}{MoNuSeg}                      & \multicolumn{2}{c|}{\begin{tabular}[c]{@{}c@{}}Gland\\ segmentation\end{tabular}} & \multicolumn{2}{l|}{DigestPath}                   \\ \cline{2-11} 
\multicolumn{1}{|c|}{}      &   MAML          &   TransferL      & MAML    & TransferL     & MAML & TransferL & MAML  & TransferL   & MAML & TransferL   \\ \hline
BreastPathQ                                        & \bf{0.301} & 0.282 & NA & NA & \bf{0.302} & 0.287& \bf{0.326}  & 0.300 & 0.285 & \bf{0.299} \\ \hline
MoNuSeg                                            & 0.669 & \bf{0.676} & \bf{0.682} & 0.636 & NA & NA & 0.691 & \bf{0.694} & 0.639 & \bf{0.653} \\ \hline
Gland segmentation 
& \bf{0.557} & 0.556   &  \bf{0.540} & 0.539 & 0.563 & \bf{0.573} & NA  & NA & 0.535 & \bf{0.553} \\ \hline
DigestPath                                         & \bf{0.632} & 0.628 & 0.609 & \bf{0.613} & \bf{0.607} & 0.599 & \bf{0.624} & 0.617 & NA & NA \\ \hline
\end{tabular}
}
\end{table}

As shown in the table, MAML and instance based transfer learning deliver comparable performance across tasks, with MAML outperforming the other in 9 of the 16 settings. 
Of note, for a number of tasks, one of the two performs noticeably better than the other. 
However, we don't observe a consistent advantage of one methodology over the other on all testing data sources. 
We hypothesize that the suitability of a knowledge sharing methodology highly depends on the interoperability between the predecessor tasks and the testing task. 
For example, when evaluating data source MoNuSeg as a new task, meta learning outperforms transfer learning with BreastPathQ as predecessor task while the reverse is true with GlandSegmentation or DigestPath as predecessor tasks. This suggests that BreastPathQ might share more task level knowledge with MoNuSeg than GlandSegmentation and DigestPath. 

\begin{figure}
\begin{centering}
\includegraphics[width=0.5\linewidth]{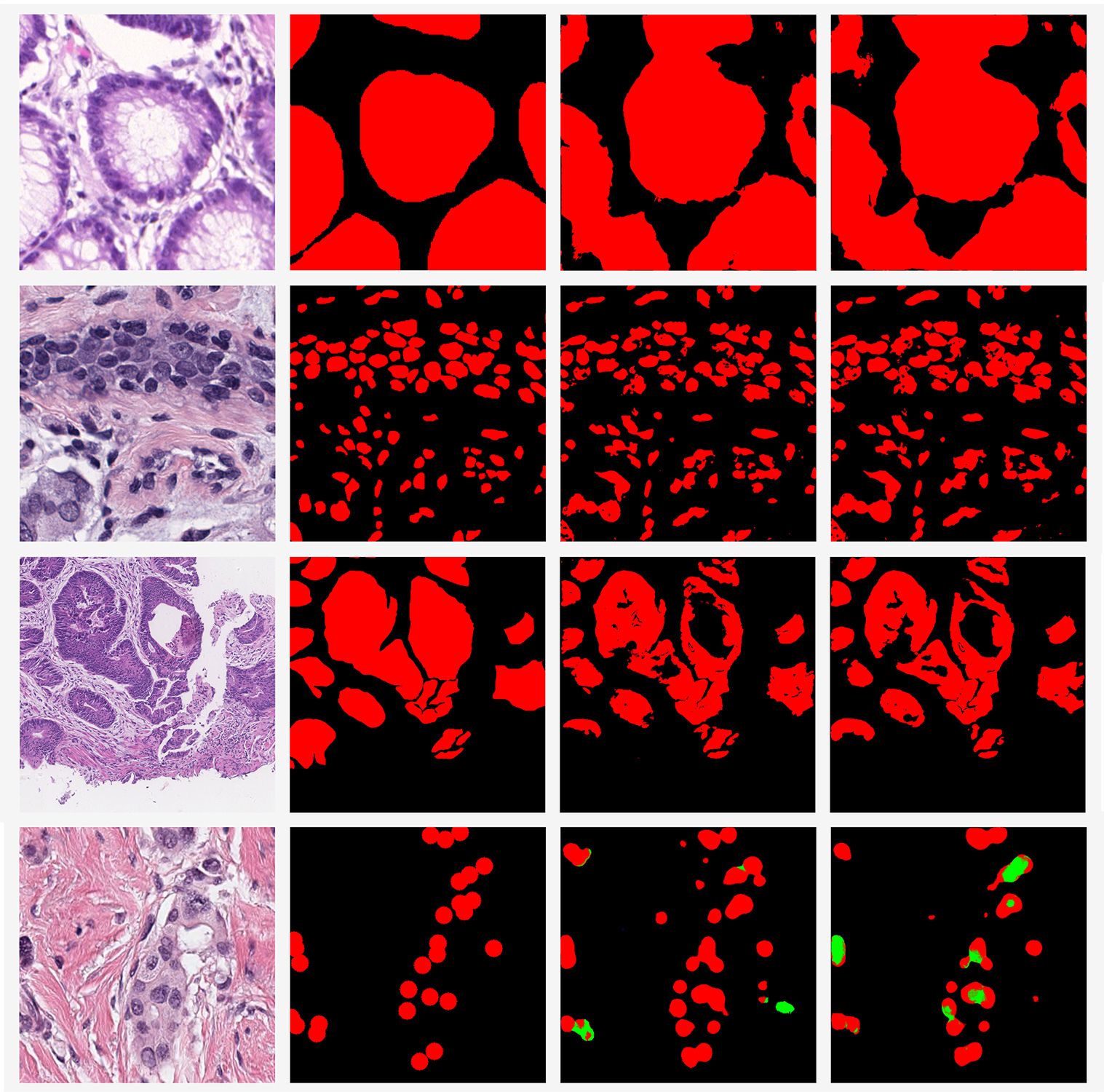}
\caption{Meta-dataset task examples. Top to bottom: GlandSegmentation, MoNuSeg, digestpath and BreastPathQ. Left to right: original image, ground truth, segmentation by MAML and segmentation by instance based transfer learning.} \label{fig:visual_results}
\end{centering}
\end{figure}

Fig.~\ref{fig:visual_results} depicts the visual comparison of two knowledge sharing methodologies. Each row is a sample of each data source while each column is original histopathology images, the ground truth masks, segmentation results from MAML and instance based transfer learning respectively. Note that for BreastPathQ (the fourth row), the raw label is standalone centroid of each nuclei and we augment them into circles with a radius of 12 pixels to generate the segmentation masks. Yet we don't impose this simplified constraint on predictions. Therefore the results are not necessarily isolated circles. We also observe that for breastPathQ, both methodologies sometimes make predictions that falsely detect (green region) a long tail class. This is due to the innate class imbalance in the data source and can be alleviated by weighted sampling. Another interesting observation is that sometimes pathologists can make ambiguous annotations. As the third row shows, there is an enclave background in the tissue while human labeler regards it as the same class as the surrounding tissue, perhaps out of medical consistency. Whereas it also makes sense in the prediction results the two methodologies still predict it as background. Overall, as shown in these figures, MAML and transfer learning produce similar qualitative results. 

\section{Conclusions}

In this work, we introduced a Python based meta learning framework MetaHistoSeg for histopathology image segmentation. Along with a curated histopathology meta-dataset, researchers can use the framework to study knowledge transferring across different histopathological segmentation tasks. 
To enable easy adoption of the framework, we provide sampling functions that realize the standard sampling procedures in classical knowledge transferring settings. We also benchmark against the meta dataset using MAML and instance based transfer learning. Based on experiment results, MAML and transfer learning deliver comparable results, and it is worthwhile to attempt each when fitting models. However, it remains unclear how interoperability of the testing task and predecessor task(s) in the training set precisely determine meta learning and transfer learning effectiveness. Also, we observe there isn't always performance gain when we add more predecessor task sources. It concludes that a naive combination of task-level training data may not be beneficial. This addressed observation points to a future research goal of explainable interoperability between tasks.

\end{document}